\documentclass[12pt]{article}

\newcommand{\bbr}{I\!\! R}
\newcommand{\bbz}{Z\!\!\! Z}

\usepackage{graphicx} 

\begin{document}
\thispagestyle{empty}
\begin{center}

\null
\vskip-1truecm
\vskip2truecm
{\bf THE STRONG ENERGY CONDITION AND THE S-BRANE SINGULARITY PROBLEM\\}
\vskip2truecm
Brett McInnes
\vskip2truecm

Department of Mathematics, National University of Singapore, 10 Kent Ridge Crescent,
Singapore 119260, Republic of Singapore.\\ 
E-mail: matmcinn@nus.edu.sg\\    

\end{center}
\vskip1truecm
\centerline{ABSTRACT}
\baselineskip=15pt
\medskip
Recently it has been argued that, because tachyonic matter satisfies the Strong Energy Condition [SEC], there is little hope of avoiding the singularities which plague S-Brane spacetimes. Meanwhile, however, Townsend and Wohlfarth have suggested an ingenious way of circumventing the SEC in such situations, and other suggestions for actually violating it in the S-Brane context have recently been proposed. Of course, the natural context for discussions of [effective or actual] violations of the SEC is the theory of asymptotically deSitter spacetimes, which tend to be less singular than ordinary FRW spacetimes. However, while violating or circumventing the SEC is necessary if singularities are to be avoided, it is not at all clear that it is sufficient. That is, we can ask: would an asymptotically deSitter S-brane spacetime be non-singular? We show that this is difficult to achieve; this result is in the spirit of the recently proved ``S-brane singularity theorem". Essentially our results suggest that circumventing or violating the SEC may not suffice to solve the S-Brane singularity problem, though we do propose two ways of avoiding this conclusion.
\vskip3.5truecm
\begin{center}

\end{center}

\newpage

\addtocounter{section}{1}
\section*{1. A Family of Asymptotically deSitter S-Brane Spacetimes}
One approach to understanding time-dependent backgrounds in string theory is based on the concept of S-branes \cite{kn:strominger}, branes with a purely spatial world-volume. Many S-brane spacetimes are known [for example, \cite{kn:gutperle}\cite{kn:peet1}\cite{kn:ohta}\cite{kn:deger}] and various interpretations of them have been proposed \cite{kn:costa1}\cite{kn:costa2}\cite{kn:wang}\cite{kn:burgess}\cite{kn:buchel1}\cite{kn:buchel2}\cite{kn:quevedo}\cite{kn:zavala}\cite{kn:costa3}. Unfortunately, S-brane spacetimes are plagued by singularities, including naked singularities. Back-reaction effects \cite{kn:peet2} tend to ameliorate these singularities, but it has been argued \cite{kn:buchel3} that they cannot be eliminated entirely. This is worrisome, since earlier investigations \cite{kn:buchel1} suggested that the naked singularities were among those resistant to resolution.

The argument of \cite{kn:buchel3} is based on a demonstration that ``tachyon matter" satisfies the Strong Energy Condition [SEC], so that the Penrose-Hawking singularity theorems guarantee the existence of some kind of singularity. Now the fact that the SEC is satisfied by string-theoretic matter in general is of course a major problem, since it is now known that the SEC is violated on a colossal scale in our Universe \cite{kn:bennett}. However, an ingenious way of dealing with this difficulty has been advanced in \cite{kn:townsend}, where it was shown that a period of cosmic acceleration can be obtained not by violating the SEC but by circumventing it by allowing the internal metric to depend on time. [One can achieve particularly good results using a hyperbolic internal space \cite{kn:silva}, as in \cite{kn:townsend}, but acceleration was obtained earlier using a time-dependent internal circle in \cite{kn:costa1}.] Admittedly, it seems that it is difficult to obtain a {\em long} period of acceleration \cite{kn:emparan}, but, in principle, it seems that the SEC problem can indeed be circumvented in this way. [See \cite{kn:neupane} for a discussion.] This is interesting, because it appears that the Townsend-Wohlfarth scenario is in fact closely related to S-Branes \cite{kn:ohta1}\cite{kn:roy}\cite{kn:ohta2}\cite{kn:emparan}\cite{kn:linde}. One can therefore hope that the TW method may not only lead to accelerated cosmologies, {\em but also to singularity resolution of S-Brane spacetimes}, by circumventing the SEC. After all, the violation of the SEC in deSitter space is responsible not just for the acceleration, but also for the absence of singularities $-$ the two features are related. [Alternatively, it has been suggested recently \cite{kn:peet3} that it may actually be possible to {\em violate} the SEC in the S-Brane context.]

While these developments once again allow us to hope that the S-Brane singularities can be resolved, circumventing or violating the SEC is not a guarantee that this will happen. Unfortunately a  fully general theory is out of reach, but we can investigate a concrete example of an asymptotically deSitter S-Brane spacetime, to see whether, in the spirit of \cite{kn:peet2}, back-reaction effects can resolve its singularities. Although this example is by no means a completely generic S-Brane spacetime, and although of course none of the putative deSitter S-brane solutions in the literature has been constructed in a fully string-theoretic context, {\em our conclusions only depend on the topology of conformal infinity}, and so they are independent of many details: they apply to the whole family of spacetimes with this asymptotic structure.  Thus, the example is much less special than it seems.

The simplest of all S-Brane spacetimes is the temporally asymptotically flat [hence non-accelerating] four-dimensional manifold obtained as follows. It is now well known that black holes in asymptotically {\em anti-deSitter space} can have topologically non-trivial event horizons \cite{kn:birmingham}; in four dimensions, with cosmological constant $-3/L^2$, such a black hole with a negatively curved [Riemann surface] event horizon has external metric 
\begin{equation} \label{eq:adshyper4}
g(external) = - {F}(r) \; dt \otimes dt + {F}^{-1}(r) \; dr \otimes dr + r^2(d\rho \otimes d\rho + sinh^2(\rho)\; d\theta \otimes d\theta)\end{equation}
with
\begin{equation} \label{eq:f4}
{F}(r) = -1 - {8\pi G \mu \over r \; {\rm{Area}}({H^2}/ {\Delta})} + {r^2 \over L^2};
\end{equation}
here $H^2$ is the hyperbolic plane of curvature -1, ${H^2}/ {\Delta}$ is a Riemann surface [with fundamental group $\Delta$], covered by $H^2$, with area ${\rm{Area}}({H^2}/ {\Delta})$ [ taking the place of the more familiar Schwarzschild unit sphere with area $4\pi$] and $\rho$ and $\theta$ are dimensionless polar coordinates on $H^2$, descending to local coordinates on $H^2/\Delta$. 
Notice that we have to factor out by $\Delta$, for two reasons: first, because the area of $H^2$ is not finite, so failing to factor by $\Delta$ would be tantamount to letting the mass tend to zero; and second, because we need a compact surface to separate the internal region from the external region $-$ otherwise we would not have a black hole. 

Now we are interested in time-dependent spacetime geometry, which, in the case of a black hole, is to be found inside the event horizon. In that region $r$ and $t$ exchange roles in the usual way, and it is better to write the metric in its non-static form as
\begin{equation} \label{eq:tr}
g(internal) = {f_{AdS}}(t) \; dr \otimes dr - ({f_{AdS}})^{-1}(t) \; dt \otimes dt + t^2(d\rho \otimes d\rho + sinh^2(\rho)\; d\theta \otimes d\theta)\end{equation}
with
\begin{equation} \label{eq:weird}
{f_{AdS}}(t) = 1 + {8\pi G \mu \over t \; {\rm{Area}}({H^2}/ {\Delta})} - {t^2 \over L^2}.
\end{equation}
[Here and henceforth we understand that the various coordinate systems we use do not cover the entire respective spacetimes. The Penrose diagrams should be used for global questions.]

There are of course well-known theorems \cite{kn:hawking} prohibiting non-spherical topology for black holes when the cosmological constant is zero or positive, so we must expect bad behaviour if we try to construct a Schwarzschild-like metric of this kind. But let us proceed anyway, letting $L$ tend to infinity [so that we obtain an asymptotically flat metric] and reversing the sign of t for convenience:
\begin{eqnarray} \label{eq:sob}
g(S0Brane) & = & (1 - {8\pi G \mu \over t \; {\rm{Area}}({H^2}/ {\Delta})}) \; dr \otimes dr - (1 - {8\pi G \mu \over t \; {\rm{Area}}({H^2}/ {\Delta})})^{-1} \; dt \otimes dt \nonumber \\ 
& & + t^2(d\rho \otimes d\rho + sinh^2(\rho)\; d\theta \otimes d\theta).
\end{eqnarray}
As the event horizon topology theorem predicts, this is not a black hole: the Penrose diagram, which takes the form given in Figure 1, shows very clearly that the singularities are naked. Generic points in this diagram correspond to Riemann surfaces. 

\begin{figure}[!h]
\centering
\includegraphics[width=0.3\textwidth]{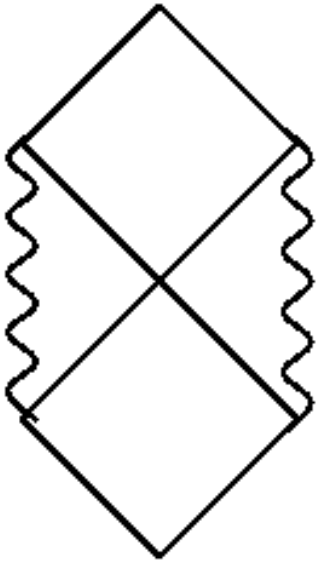}
\caption{Penrose diagram of g(S0Brane)}
\end{figure}

The metric we have found is in fact precisely the metric corresponding to an S0-Brane \cite{kn:wang}\cite{kn:burgess}\cite{kn:buchel1}, and the naked singularities shown are the subject of the recent work \cite{kn:peet2}\cite{kn:buchel3}\cite{kn:peet3} with which we are concerned here. The point of this derivation of the S0-Brane metric from the AdS black hole metric is that it is now easy to derive the metric in the case where the geometry is asymptotically deSitter instead of asymptotically flat. But, in addition, the derivation makes it clear that the naked singularities arise from the deep structure of space-time geometry $-$ the topological restrictions on black hole event horizons. They are not accidental results of choosing to examine a particular, relatively simple spacetime.

Turning to the deSitter case, the event horizon topology theorem again guarantees trouble if we reverse the signs of $1/L^2$ [to obtain a positive cosmological constant] and $t$ in equation (4). Doing so, we obtain the ``deSitter S0-Brane" metric
\begin{equation} \label{eq:t}
g(deS0Brane) = {f_{dS}}(t) \; dr \otimes dr - ({f_{dS}})^{-1}(t) \; dt \otimes dt + t^2(d\rho \otimes d\rho + sinh^2(\rho)\; d\theta \otimes d\theta)\end{equation}
with
\begin{equation} \label{eq:weir}
{f_{dS}}(t) = 1 - {8\pi G \mu \over t \; {\rm{Area}}({H^2}/ {\Delta})} + {t^2 \over L^2}.
\end{equation}
[Again, if this looks odd, the reader should remember that the coordinates cover only part of the spacetime. In this case we are interested in very late times, and these coordinates are valid then.] The Penrose diagram in this case is just like that in Figure 1, but, since the spacetime is asymptotically deSitter, the conformal boundary is spacelike instead of null. [This  metric was discussed in a different context in \cite{kn:cai} and in an appendix of \cite{kn:leblond}.] Again, of course, the  singularities are naked.

\begin{figure}[!h]
\centering
\includegraphics[width=0.3\textwidth]{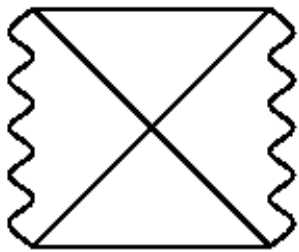}
\caption{Penrose diagram of g(deS0Brane)}
\end{figure}

Our question is this: if we introduce some matter into a spacetime of this kind, and take the back-reaction into account, can the geometry be deformed so that the singularities are resolved? The point is that, with the aid of explicit SEC-violation resulting from the positive cosmological constant in the background of $g(deS0Brane)$, it should be easier to achieve this here than it is in the asymptotically flat case. Certainly the singularity theorem of \cite{kn:buchel3} will not apply.

Our derivation suggests, however, that it will be no easy matter to resolve the naked singularities, {\em even with the aid of a SEC-violating positive cosmological constant}. For we have seen that the trouble arises from the {\em topology} of the spatial sections, and it is hard to see how this can be changed. As we shall see in the next section, there are other pieces of evidence suggesting that it will be difficult to eliminate the singularities of this spacetime.

\addtocounter{section}{1}
\section*{2. Positive Scalar Curvature at Infinity}
In an important work motivated by the dS/CFT correspondence, Andersson and Galloway \cite{kn:galloway} have given a singularity theory for asymptotically deSitter spacetimes. 
In particular they give a rigorous definition of the term ``asymptotically deSitter" in terms of manifolds with boundary, in an analogous way to the usual definition of asymptotic flatness; see \cite{kn:galloway} for the details, and note that it is {\em not} assumed a priori that the boundary components have 3-sphere topology [as is of course the case for deSitter space itself.] A spacetime of this kind is said to be {\em future [past] asymptotically simple} if every future [past] inextendible null geodesic has an endpoint on future [past] conformal infinity. Then Andersson and Galloway prove the following remarkable result. Suppose that an asymptotically deSitter spacetime is future or past asymptotically simple and that it satisfies the {\em Null} [not the Strong] Energy Condition. Then the spacetime must be globally hyperbolic, with compact Cauchy surfaces having a finite fundamental group. This also implies that future and past conformal infinity are compact with finite fundamental group. Essentially this just means that an asymptotically deSitter spacetime which is non-singular must either violate the Null Energy Condition or have conformal infinities which, as in deSitter space, are compact with finite fundamental group.

The relevance of this result here becomes clear if we allow $t$ to tend to infinity in equations   (6) and (7). The result is a representative of the conformal structure at future infinity:
\begin{equation} \label{eq:t}
{g(deS0Brane)}^{+\infty} = dr \otimes dr + L^2(d\rho \otimes d\rho + sinh^2(\rho)\; d\theta \otimes d\theta).\end{equation}
Since the topology of a Riemann surface is ${\bbr}^2/\Delta$, this is a metric either on $\bbr \times ({\bbr}^2/\Delta)$ if $r$ is interpreted in the most obvious way [that is, if the two naked singularities are distinct] or on $S^1 \times ({\bbr}^2/\Delta)$ if $r$ is identified periodically. This second interpretation is actually rather reasonable, since there seems to be no physical reason to assume that S-Branes must always occur between symmetrical pairs of singularities. That is, in this picture, there is only one naked singularity, and the appearance of two in Figure 2 is due to ``unwrapping" $S^1$. [This is analogous to the interpretation of Schwarzschild-deSitter space in which the topology of the spatial sections is $S^1 \times S^2$.] Compactifying in this way is also suggested by the Andersson-Galloway theorem, which requires a compact conformal infinity. With either interpretation, however, we can see that {\em the topology of infinity in this spacetime actually forbids any resolution of the singularity}. For neither $\bbr \times ({\bbr}^2/\Delta)$ nor $S^1 \times ({\bbr}^2/\Delta)$ has a finite fundamental group: every compact Riemann surface has an infinite fundamental group. The Andersson-Galloway theorem therefore implies that even if we allow matter to back-react on the geometry of this space-time [after the manner of \cite{kn:peet2}] then, as long as the Null Energy Condition continues to hold, the spacetime can never be asymptotically simple. [Taking the universal cover of the Riemann surface is not an option here, because the theorem also requires compactness, and the universal cover of a Riemann surface is of course non-compact.] Some inextendible null geodesic fails to arrive at infinity: it is waylaid by a singularity. The key point here is that the Andersson-Galloway theorem is a result on the {\em topology} of infinity, and it is hard to believe that back-reaction effects can effect a [necessarily discontinuous] modification of that topology. Thus it appears that we cannot resolve the S-Brane singularities merely by circumventing or violating the SEC, {\em unless} we are willing to go beyond that and violate the Null Energy Condition. This latter point will be considered in the next section.

We can understand this situation from a different, perhaps more physical point of view as follows. The metric (8) has scalar curvature equal to $-2/L^2$. Experience with the AdS/CFT correspondence has taught us \cite{kn:witten1}\cite{kn:witten2}\cite{kn:witten3}\cite{kn:witten4} that it is dangerous to allow the scalar curvature of the conformal boundary to be negative, particularly when that boundary is Euclidean, as is the case here. One way of explaining the problem is simply to note that the conformal Laplacian [in three dimensions] is
\begin{equation} 
{\nabla}^2_{conformal} = {\nabla}^2 + {1 \over 8}R,
\end{equation}
where $R$ is the scalar curvature. This is not a non-negative operator if $R<0$, and this leads to instability [``Seiberg-Witten instability" \cite{kn:witten4}] in any ``holographic" CFT on a boundary of negative scalar curvature. Now even if the reader does not believe in a dS/CFT correspondence \cite{kn:strominger1}\cite{kn:strominger2}, and doubts have been raised \cite{kn:susskind}, it is a fact \cite{kn:staruszkiewicz} that the limiting values of correlation functions for a theory on a deSitter background do define correlation functions for a Euclidean CFT on the boundary. Hence, Seiberg-Witten-style misbehaviour of a CFT on a boundary manifold with a metric like (8) surely indicates some kind of instability in the bulk, perhaps resulting in a singularity, whether or not a full ``correspondence" [in the AdS/CFT sense of complete equivalence] is valid.

Now, once again, we can ask whether back-reaction of matter introduced into the spacetime can modify the geometry at infinity in such a way that the scalar curvature there is in fact positive, thereby averting Seiberg-Witten instability. This would be analogous to the mechanism of singularity resolution by back-reaction discussed in \cite{kn:peet2}. However, it is actually possible to prove $-$ see \cite{kn:mcinnes1} for a discussion of the mathematical techniques $-$ that, no matter how $S^1 \times ({\bbr}^2/\Delta)$ is distorted, {\em it is completely impossible to force its scalar curvature to be positive [or zero] everywhere}. [Surprisingly, this is also true of geodesically complete metrics on $\bbr \times ({\bbr}^2/\Delta)$; the non-compactness does not help in this particular case.] Thus, back-reaction cannot prevent the instability from getting out of control. This is an example of ``topologically induced instability" \cite{kn:mcinnes2}: once again, it is the topology of the boundary that enforces the instability, not its geometry. Thus, from this [apparently] quite different point of view, we again see that back-reaction is ineffective in resolving the singularity of the deSitter S-Brane spacetime.

{\em Every} Riemannian metric on $S^1 \times ({\bbr}^2/\Delta)$, and every geodesically complete metric on $\bbr \times ({\bbr}^2/\Delta)$, has a scalar curvature function which is negative somewhere on the manifold. Consider any four-dimensional asymptotically deSitter spacetime with compact conformal infinity having this property. Then [modulo standard conjectures on three-manifold topology, namely the Poincar\`e conjecture and the conjecture that all free actions by finite groups on $S^3$ are orthogonal] the fundamental group {\em must} be infinite; see \cite{kn:schoen}. As we have seen, this forces the spacetime to be singular. That is, in four dimensions, the Andersson-Galloway singularity and the Seiberg-Witten instability have a common origin in a curious property of certain manifolds: their refusal to accept any metric of positive scalar curvature. [This fascinating and extremely deep branch of geometry has been applied recently \cite{kn:horowitz} to Calabi-Yau manifolds.] This suggests a second way out of the problem, which we shall discuss in the next section.

\addtocounter{section}{3}
\section*{3. Two Ways Out}
The Andersson-Galloway theorem, and our discussion of topologically induced instability, suggest two ways out of the unwelcome conclusions of the preceding section. Either we must violate the Null Energy Condition, or we must do something about the topology of conformal infinity.

While it is almost certain that the Strong Energy Condition is violated in our Universe, the Null Energy Condition [NEC] is a much weaker condition, and it {\em may} hold true. The {\em theoretical} evidence in favour of it is not as strong as is sometimes thought; the ``traditional" arguments in favour [such as the fear that the speed of sound might exceed that of light] are weak indeed $-$ see \cite{kn:mcinnes3} for a review of these and more recent arguments for and against the NEC. The {\em most obvious} kinds of NEC violations do seem to lead to serious problems \cite{kn:carroll}\cite{kn:gibbons}, but the theoretical situation is unclear for more complex models; see \cite{kn:fabris}\cite{kn:hao1}\cite{kn:hao2}\cite{kn:nojiri1}\cite{kn:nojiri2}\cite{kn:dadhich} for recent theoretical developments, and note particularly that NEC violation arises rather naturally in string-theoretic accounts of ``bounce" cosmologies \cite{kn:kachru}. 

In cosmology, a violation of the NEC would correspond to a value for $w$, the dark energy equation-of-state parameter, below -1. Obviously the real answer to the question, ``Can the dark energy equation-of-state parameter w be less than -1?" must come from observations, not theory. In fact the latest data offer very little support for a value above -1; if anything, they [slightly] favour $w<-1$; readers are urged to examine figures 13 and 14 of \cite{kn:filippenko}. The relevant WMAP data are discussed in \cite{kn:odman}\cite{kn:caldwell}. It is clear from these references that the possibility of NEC violation in cosmology cannot be ignored.

In view of this, we should ask whether the assumption that the NEC holds in the Andersson-Galloway theorem is really necessary. The answer is that it is indeed necessary: in \cite{kn:mcinnes4} a family of cosmological models was presented with the metric 
\begin{equation}
g(\gamma ,A) = -dt \otimes dt + A^2 {\cosh}^{\left({4\over \gamma}\right)}
	\left( {\gamma t \over 2L}\right) (d\theta_1 \otimes d\theta_1 + 
d\theta_2 \otimes d\theta_2 + d\theta_3 \otimes d\theta_3). 
\end{equation}
Here $A$, $L$, and $\gamma$ are positive constants, and the spatial sections are cubic tori with angular coordinates $\theta_i$. This spacetime is asymptotically deSitter and future and past asymptotically simple, but the Cauchy surfaces [and future and past conformal infinity] are tori with infinite fundamental groups. The apparent contradiction with the Andersson-Galloway theorem is explained when one examines the NEC in this spacetime: it is indeed violated by a time-dependent $w$ which is always less than -1, but which approaches -1 rapidly towards the future and past. [This last property allows the spacetime to avoid the ``Big Smash" \cite{kn:mcinnes4} or ``Big Rip'' \cite{kn:caldwell} which is the fate of NEC-violating cosmologies with {\em constant} $w$.] Hence our conclusions based on the Andersson-Galloway theorem in the previous section did indeed depend on the validity of the NEC. One possible way out of the S-Brane singularity problem is therefore to arrange for the NEC to be either circumvented [in the Townsend-Wohlfarth manner] or explicitly violated [perhaps ``effectively", as in the Kachru-McAllister \cite{kn:kachru} cosmology]. Constructing an explicit singularity-free model of this kind may however be a challenging exercise; we suggest that the metric $g(\gamma ,A)$ may be a good place to start.

An alternative approach would proceed as follows. We saw above that the unpleasant features of the deSitter S0-Brane spacetime arose, fundamentally, from the topology of conformal infinity. Can Figure 2 be re-interpreted in such a way that the topology is usefully different? As we saw, one can question whether the ``two" singularities are really distinct; perhaps they should be identified. The topology of future infinity is then $S^1 \times ({\bbr}^2/\Delta)$. That certainly does not help to render the fundamental group more finite, but it does suggest the following approach. Suppose that we perform a $\bbz_2$ projection on the circle, in the manner of the Horava-Witten \cite{kn:horava} model, so that the singularity is at one fixed point. Then the topology of the future and past infinities of our spacetime becomes ${{S^1}\over{\bbz_2}} \times ({\bbr}^2/\Delta)$. The Penrose diagram is as shown in Figure 3.

\begin{figure}[!h]
\centering
\includegraphics[width=0.3\textwidth]{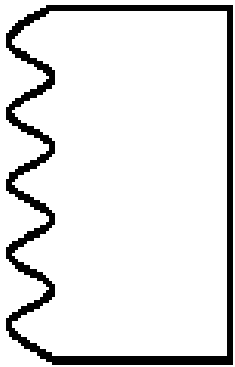}
\caption{As in figure 2, but factored by $\bbz_2$}
\end{figure}

Here the vertical line on the right represents the other fixed point on the circle; the reader should think of it as a timelike brane, just as in brane cosmology. At infinity, instead of a conformal manifold, we have a conformal manifold with boundary. This may not seem to be much of a difference, but in fact it changes everything. As we stated above, it is completely impossible to find a metric of positive scalar curvature on either $\bbr \times ({\bbr}^2/\Delta)$ or $S^1 \times ({\bbr}^2/\Delta)$. But it {\em is} possible to find a metric of positive scalar curvature on ${{S^1}\over{\bbz_2}} \times ({\bbr}^2/\Delta)$. A simple example is as follows. Let $r$ parametrise the closed interval $[-{{L}\over {2}}, +{{L}\over {2}}]$, and let $\rho$ and $\theta$ be local coordinates on ${\bbr}^2/\Delta$ as in equation (8). Then [recalling that $S^1/\bbz_2$ is the same as a closed line interval] we can define a metric on ${{S^1}\over{\bbz_2}} \times ({\bbr}^2/\Delta)$ by
\begin{equation} 
g = sech^2({{r}\over {L}})[ dr \otimes dr + L^2(d\rho \otimes d\rho + sinh^2(\rho)\; d\theta \otimes d\theta)].
\end{equation}
The scalar curvature of this metric is simply
\begin{equation} 
R(g) = {{2}\over {L^2}}[3 \: - \: 2cosh^2({{r}\over {L}})],
\end{equation}
which is indeed positive everywhere since $r$ is confined to $[-{{L}\over {2}}, +{{L}\over {2}}]$. This example is too simple to be useful, since it is conformal to a metric of negative scalar curvature, but it does illustrate the point that curvature controls topology much more weakly for manifolds with boundaries than it does for ordinary manifolds. [It also shows why this is so: we can use the boundary conditions to manipulate the parameters. For example, simply by choosing a different range for $r$ $-$ say [L, 2L] $-$ we can force the scalar curvature to be {\em negative} everywhere on this same manifold with boundary, with a metric defined by the same formula.]

In fact, Gromov \cite{kn:gromov} has proved that it is possible to find a metric of positive {\em sectional} curvature] on {\em every} manifold with boundary [of dimension at least two]. It is also possible to find a metric of negative sectional curvature on the same space. That is, essentially, {\em every} curvature component can be made positive or negative, not just the sum. In effect, this means that there is no relation between topology and curvature on manifolds with boundaries. [This is of course in very sharp contrast to the case of ordinary manifolds: in particular, in three dimensions, a compact manifold with positive [not necessarily constant] sectional curvature has to have the topology of a sphere or of a finite quotient of a sphere.] Therefore, to the extent that our problems are due to the absence of positive scalar curvature metrics on the conformal boundary, we can in principle solve them in this way. 

We do not know whether the Andersson-Galloway theorem can be extended to cover this case. Let us assume the worst and suppose that it can. Even then, we have to re-consider our assumption that a failure of asymptotic simplicity signals the presence of a singularity. For inextendible null geodesics will have endpoints on the timelike brane in Figure 3 $-$ and therefore fail to reach conformal infinity $-$ whether or not the naked singularity remains. That is, the spacetime can fail to be asymptotically simple in this case without being singular. Once again, the obstruction to resolving the naked singularity has been removed: in effect, we can try to replace one kind of ``discontinuity" by another which is less obnoxious [because it is under control, as a physical object]. 

To summarise: $g(deS0Brane)$ is undoubtedly a metric on a nakedly singular spacetime. But if we think of it as a metric on the space pictured in Figure 3, with future and past boundaries having the structure ${{S^1}\over{\bbz_2}} \times ({\bbr}^2/\Delta)$, then there is nothing to prevent back-reaction from changing the sign of the scalar curvature at infinity, and there is no reason to insist that the failure of some null geodesics to reach infinity signals the presence of a singularity. That might well mean that the singularity in the original spacetime can be resolved. As in the case of the first escape route, however, it will not be easy to produce a concrete example.

In view of the Andersson-Galloway theorem and the topological nature of Seiberg-Witten instability in this case, it is hard to see any other possible way of resolving the singularities in evidence in Figure 2. If the reader feels that neither method proposed in this section is likely to work, then he or she can conclude that violating the SEC is unlikely to change the conclusions of \cite{kn:buchel3} regarding the intrinsically singular nature of this whole family of spacetimes [that is, those with asymptotic structure of the kind typified by equation (8).] It seems that we need a different ansatz, one which not only does away with the symmetries [as proposed in \cite{kn:buchel3}] but also has a different underlying topology.

\end{document}